# A New Fuzzy DEMATEL-TODIM Hybrid Method for evaluation criteria of Knowledge management in supply chain

Mahdi Mahmoodi, MS student, Industrial engineering Department, Kharazmi University of Tehran,
Gelayol Safavi Jahromi, MS student, MBA information systems, Management and economics department, Semnan University,

## ABSTRACT:

*Knowledge management (KM) adoption in the supply chain network needs a good investment as well as few changes in the culture of the entire SC. Knowledge management is the process of creating, distributing and transferring information. The goal of this study is to Rank KM criteria in supply chain network in Iran which is important for firms these days. Criterion used in this paper were extracted from the literature review and were confirmed by supply chain experts. The proposed approach for ranking and finding out about these criterion is hybrid fuzzy DEMATEL-TODIM, with using fuzzy number as data for our studies we could avoid uncertainty. The data was gathered from PhD. And Ms. Students in industrial engineering of Kharrazmi university of Tehran and PhD. And Ms. Students of the management department of Semnan university. A new hybrid approach was used for achieving the results of this study. This new hybrid approach ranks data criteria respect to each other, then by using TODIM for ranking respect to the best situation (gains), the rates of criterion were determined which is a very important advantage.*

## 1. Introduction

Knowledge management (KM) is a field of study that has attracted the attention of many of researchers in last years. The main focus of KM research to date has been on large organization processes in order to improve their performance and competitive position, assuming that those organizations have required resources, Edvardsson & Durst, (2013)[1]. The goal of Knowledge Management Systems (KMS) is supporting creation, transfer and application of knowledge in organizations salami et al. (2012)[2].

Supply chain management has received a great deal of attention by practitioners and academics in recent years. The benefit of effectively managing supply chain partners for organizations, range from lower costs to higher return on investment (ROI), to higher returns to stockholders ,spekman et al, (2002)[3].

KM is recognized as an important source of competitive advantage and hence there has been increasing academic and practitioner interest in understanding and isolating the factors that contribute to effective knowledge transfer between Supply Chain (SC) actors. The KM adoption in SC, makes the environment collaborative that enables the chain to be more adaptive and responsive. This helps the organization to achieve an improved strategic competitive position in the market. KM among SC members can provide a guarantee that chain members can access the external knowledge, and also it is helpful to make the overall environment of the entire SC more competitiveness. Generally, managing knowledge within SC can help organizations to have better use of resources. KM and SC represent two main fields of research that have developed over the past years. But many related issues are still not addressed by consultants, practitioners or academics yet. ,Patil & Kant, (2013)[4].

According to Zyl (2003)[5], the adoption of KM practices and principles, and the subsequent automation of the supply chain through collaborative knowledge portals and electronic document management systems, enables organizations within the supply chain to develop a more cost-effective, efficient and competitively responsive and adaptive supply chain. However, this can only be achieved if the guidelines for successful supply chain KM implementation, and subsequent use and adoption, are seriously addressed and followed.

In this research, researchers have tried to extract criterion of KM approach in supply chain from the literature review. These criterion has been determined respect to the needs of supply chain in Iran. For determining the weights of criterion, new fuzzy hybrid DEMATEL-TODIM has been used which was introduced for the first time in this research. The advantage of this new approach is that it gives some feedbacks about the criterion and the gap. Using fuzzy triangular number in this combination also helped to avoid uncertainty in this research.

## 2. Literature review

There have been lots of researches about KM adoption in supply chain. In a research done by Maqsood and Walker (2007)[6], through extensive literature review, commonalities between knowledge management and supply chain management were elicited. Knowledge Advantage framework, which was developed as a part of CRC for Construction Innovation Australia, research project “Delivering improved knowledge management and ICT diffusion in Australian construction industry”, has been proposed to extend across the supply chain in order to develop learning chains. The paper suggests that, as unit of competition changes from organization vs. organization to chain vs. chain under supply chain management, learning organization can not answer to the complex and dynamic business environment by itself. The learning chains are to be created instead, through managing knowledge in supply chains. This will facilitate innovation and creativity required to thrive in today unpredictable business environment.

Spekman et al. (2002)[3], claim that they explored the pre-conditions for learning to emerge and the impact of learning on supply chain performance. They introduced two measures of the relative magnitude to which supply chains encourage and/or provide support to the learning processes. One of them was designed to reflect the extent to which attitudes and behaviors within supply chains encourage the processes of learning. Items reflected the extent to which the supply chain encouraged idea sharing and supported experimentation. The second measure reflected the extent to which there are supply chain structures that support a learning environment. Here items reflected the extent to which systems and structures supported the generation of idea and sharing those ideas across the supply chain. The items were first selected based upon faced validity, and reliability was assessed using cronbach’s alpha. They assessed the relationship between pre-conditions for learning and learning in addition to the relationship with these variables and measures of performance using ordinary least squares regression procedures. Finally they have lent credibility to the learning importance to developing effective enterprise-wide SCM and have linked it to measures of performance.

In another research done by yang (2012)[7], his goal was to investigate how different knowledge-management processes (i.e. knowledge acquisition and dissemination) affect the manufacturers' performance in collaborative economic exchanges with their suppliers. He introduced some measures in different categories like knowledge acquisition, knowledge dissemination, supply-chain integration, relational stability and alliance performance. Having knowledge based view and according to transaction cost economics, this study proposes that knowledge-management processes are positively related to the performance of the manufacturers in a collaborative buyer–supplier relationship. It can be claimed that when the levels of supply chain integration and relational stability are higher rather than lower, this link is stronger. The findings of this study show strong support for these propositions.

In the research done by Patil and Kant (2014)[8], they first identified the evaluation criteria of KM adoption in SC from literature review and expert opinion. Further, they used fuzzy DEMATEL to evaluate weighting of each evaluation criteria's, after that FMCDM method used to obtain possible rating of success of KM adoption in SC. The proposed approach is helpful to predict the success of KM adoption in SC without actually adopted KM in SC. It also enables organizations to decide whether to initiate KM, restrain adoption or undertake remedial improvements to increase the possibility of successful KM adoption in SC.

Recently researchers are using MCDM tools for evaluating weights of criterion. Gomes et al.(2009) [9] used TODIM to investigate and recommend options for upstream projects for the natural gas reserves that were recently discovered in the Brazil. In addition, Gomes and Rangel by using the TODIM method of multi criteria decision could present an evaluation of residential properties with real estate agents in Brazil and defined a reference value for the rents of these properties' characteristics. Adil Baykasoglu et al.(2013) used fuzzy DEMATEL AND fuzzy hierarchical TOPSIS methods for truck selection ,the proposed approach was new hybrid of DEMATEL and TOPSIS .In this research we used combination of DEMATEL and TODIM to evaluate criteria weights of KM in supply chain networks .

# 3. Methodology

By reviewing the literature, we found lots of indices for KM adoption in SCM but after consulting with some SC experts, we reduced them into 17 indices as shown in table 1.

Table 1: important indices in adopting KM into SC.

| Category | Indices | References |
|---|---|---|
| Employee trait | 1- Virtual teaming<br>2- Capacity to develop knowledge within SC<br>3- Employee involvement | Patil and Kant (2014) |
| Re Organizational strategy | 4- Developing new insights<br>5- Supporting experimentation in the SC<br>6- Giving reward to employees for their new ideas in the supply chain | Spekman et al. (2002)<br>Maqsood and Walker (2007) |
| Organizational culture | 7- problem-solving in a systematic way<br>8- Learning from their own experiences and past activities<br>9- Learning from the experiences and best practices of others<br>10- quick and efficient transferring of knowledge throughout the organization | Maqsood and Walker (2007) |
| Technological factors | 11- Electronic data interchange (EDI) links<br>12- IT integration with all suppliers/customers<br>13- Techniques of networking<br>14- Integrated business systems | Spekman et al. (2002)<br>Patil and Kant (2014) |
| Managerial factors | 15- Establishing more frequent contact with supply-chain members<br>16- Creating a compatible communication/information system<br>17- Employee empowerment | Yang (2014)<br>Patil and Kant (2014) |

Indices that were introduced in table 1, were extracted from literature review and the correlation between KM in SC with categories and items were cited in papers of those researchers introduced in the third column of the table. For example in a research done by Kant and Patil (2014), they have used some methods to predict the success of KM adaption in SC. In this research, they had a table summarizing the literature review and expert opinion on evaluation criterion of KM adaption in SC. Some of the indices in table 1 were extracted from their research. This is the same about the rest of the indices cited in table 1.

According to table 1, the framework of this research is like below:

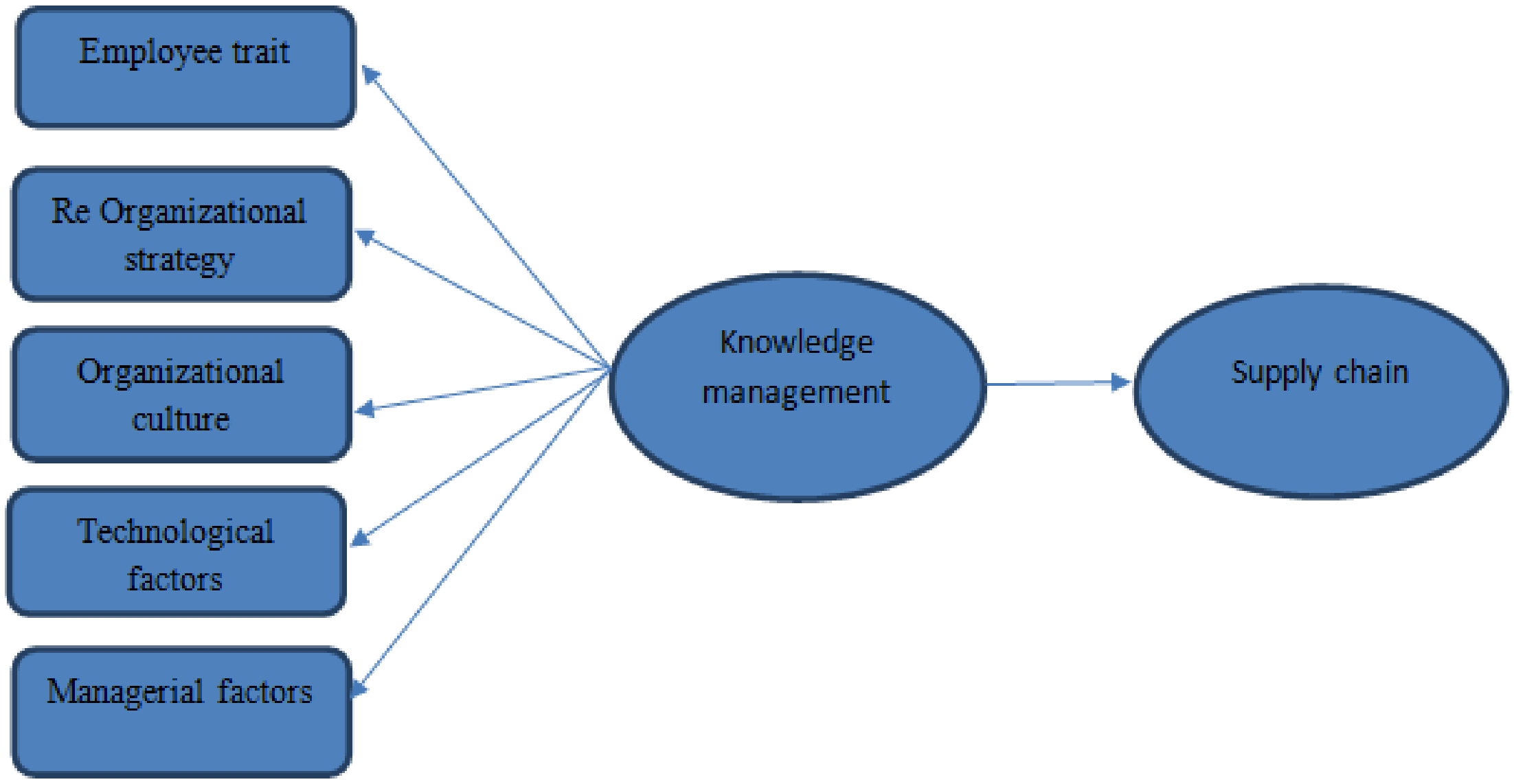

Figure 1.The hypothesis of the research is that criterion that were introduced in table 1, are key criterion of KM adaption in supply chain.

## 3.1. Fuzzy Logic

A fuzzy set is a class of objects with grades of membership. A membership function for fuzzy numbers is between zero and one Zadeh, (1965)[11]. Fuzzy logic is derived from fuzzy set theory to deal with reasoning that is approximate rather than precise, also using fuzzy logic helps to avoid uncertainty . It allows the model to easily incorporate various subject experts' advice in developing critical parameter estimates, Zimmermann, (2001)[12]. In other words, fuzzy Logic enables researchers to handle uncertainty.

There are some kinds of fuzzy numbers. Among the various shapes of fuzzy number, triangular fuzzy number (TFN) is the most popular one. It is represented with three points as A = (a1, a2, a3). The membership function is illustrated in (1). Let A and B are defined as A = ($a_1$, $a_2$, $a_3$), B = ($b_1$, $b_2$, $b_3$). Then C = ($a_1 + b_1$, $a_2 + b_2$, $a_3 + b_3$) is sum of these two numbers. Besides, D = ($a_1 - b_3$, $a_2 - b_2$, $a_3 - b_1$) is the subtraction of them. Figure 1 shows the triangular fuzzy numbers.

$$\mu_{\tilde{a}} = \begin{cases} 0, & x < a_1 \\ \dfrac{x - a_1}{a_2 - a_1} & a_2 \geq x \geq a_1 \\ \dfrac{a_3 - x}{a_3 - a_2} & a_2 \geq x \geq a_3 \\ 0, & otherwise \end{cases} \qquad Eq.(1)$$

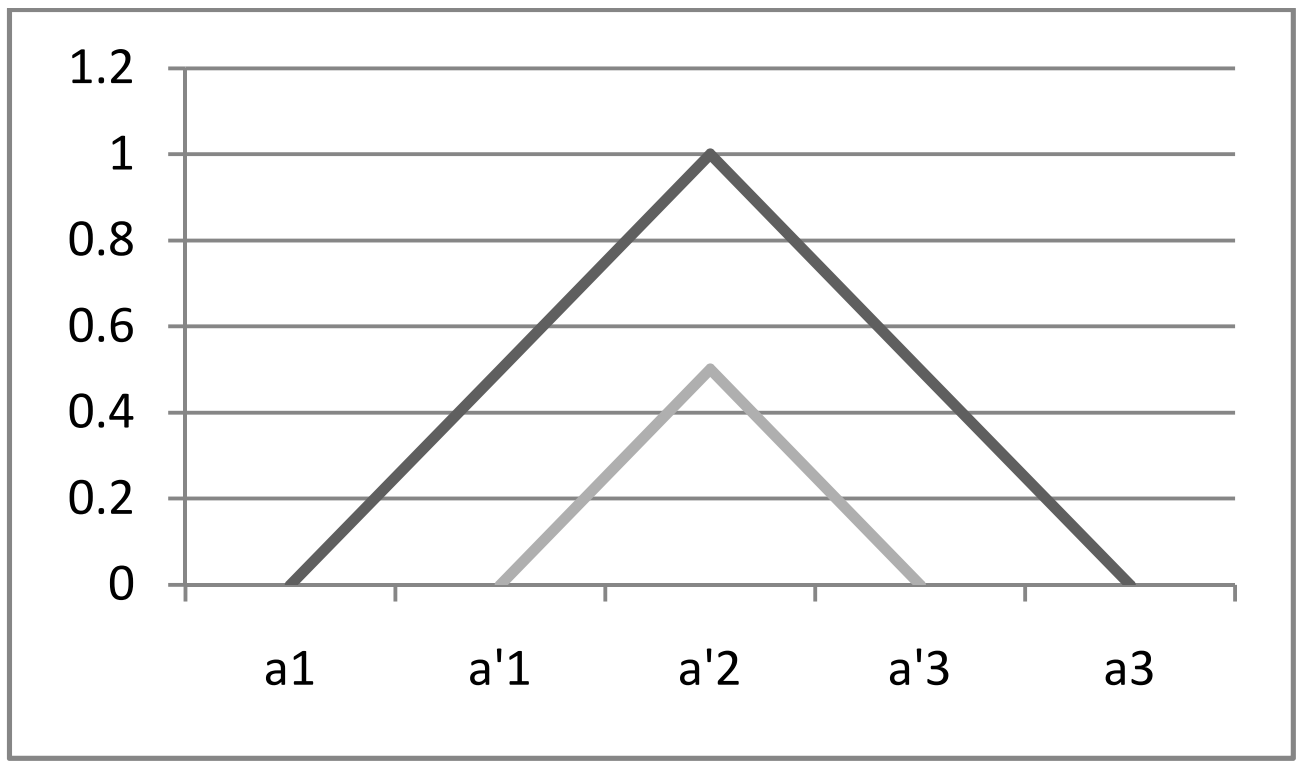

Figure 2.Traingular fuzzy number

## 3.2. DEMATEL

The common relationship between criterion is determined by using fuzzy DEMAEL method. The DEMATEL method originated from Geneva Research center of the Battelle Memorial institute, is especially pragmatic to visualize the structure of complicated casual relationships, Buyukozkan & Cifici,2011.[13]fuzzy DEMATEL and fuzzy hierarchical TOPSIS methods used for truck selection for solving transportation issue in company .Adil Baykasoglu, et al.(2013)[14].The original DEMATEL method is modified by some researchers so as to make it comply with their problems. In the modified DEMATEL model of Dalalah et al (2011)[15] the relationship between criterion are represented with direct relation matrix, in this paper we used combination of direct-indirect relation matrix to determine initial weights of criterion.

Since the form of fuzzy numbers is not suitable for matrix operation, defuzzfication algorithm is needed for further aggregation. In previous researches, Euclidean distance has been using in TODIM and DEMATEL for defuzzification step. In this study, researchers have used a method by which they converted fuzzy number into a crisp number which is called best non-fuzzy performance value (BNP). This paper employs CFCS (Converting fuzzy data into crisp scores) for defuzzification. The fuzzy aggression procedure can be shown as follow:

**Step1**: i. Standardization of fuzzy numbers

$$TFN(l,m,r), l \le m \le r$$
$$k, number\ of\ \exp erts$$

$$xl_{ij}^{k} = \frac{l_{ij}^{k} - \min\limits_{1\le k\le K} l_{ij}^{k}}{\otimes \min\max} \qquad (2)$$

$$xm_{ij}^{k} = \frac{m_{ij}^{k} - \min\limits_{1\le k\le K} l_{ij}^{k}}{\otimes \min\max} \qquad (3)$$

$$xr_{ij}^{k} = \frac{r_{ij}^{k} - \min\limits_{1\le k\le K} l_{ij}^{k}}{\otimes \min\max} \qquad (4)$$

$$where\ \otimes \min\max = \max r_{ij}^{k} - \min r_{ij}^{k}$$

ii. Calculate the left and right normalized value

$$xls_{ij}^{k} = \frac{xm_{ij}^{k}}{1 + xm_{ij}^{k} - xl_{ij}^{k}} \qquad (5)$$

$$xrs_{ij}^{k} = \frac{xr_{ij}^{k}}{1 + xr_{ij}^{k} - xm_{ij}^{k}} \qquad (6)$$

iii. Compute the total normalized value

$$x_{ij}^{k} = \frac{xls_{ij}^{k}(1 - xls_{ij}^{k}) + xrs_{ij}^{k}.xrs_{ij}^{k}}{1 + xrs_{ij}^{k} - xls_{ij}^{k}} \qquad (7)$$

iv. Compute crisp value

$$BNP_{ij}^{k} = \min l_{ij}^{k} + x_{ij}^{k} \otimes \min\max \qquad (8)$$

V. integrating crisp value

$$a_{ij} = \frac{1}{k} \sum_{k}^{1\le k\le K} BNP_{ij}^{k} \qquad (9)$$

With this approach the 'A' matrix will be determined. 'A' matrix is the initial matrix for DEMATEL approach which is used to calculate weights of criterion. Then the weight of criterion has been used for initial weightings in TODIM approach which helps to hybrid this to approach and use both to final weightings of criterion in KM for supply chain management. The TODIM approach helps to find out more specifically the difference between criteria and the gap between the current situation and the desired situation.

$$A = \begin{pmatrix} x_{11} & \cdots & x_{1n} \\ \vdots & \ddots & \vdots \\ x_{m1} & \cdots & x_{mn} \end{pmatrix} \qquad (10)$$

Step 2:In the next step, where A is n*n, criteria respect to criteria, non-negative matrix, aij shows the direct impact of factor I on factor j; and when i=j, the diagonal elements aij=0.

Step 3: calculating the normalized direct-relation matrix D=[dij] , which can be obtained through (11).

$$D = \frac{1}{\max\limits_{1 \le i \le n} \sum\limits_{j=1}^{n} a_{ij}} A \quad (11)$$

Step4: Calculating the total relation matrix T by using (12) in which L is an n*n identity matrix. Theelements tij indicates the indirect effects that factor I has on factor j, so the matrix T can reflect the total relationship between each pair of system factors.

$$T = D(1 - D)^{-1} \qquad (12)$$

**Step 5**: **initial weighting**: To make outcome more visible, $r_i$ and $c_j$were compared through (11) and (12), respectively. The sum of row I, which is denoted as $r_i$ , represents all direct and indirect influences given by factor I to all other factors, and so $r_i$ can be called the degree of influential impact. Similarly, the sum of column j, which is denoted as $c_j$ can be called as the degree of influenced impact, since $c_j$ summarizes both direct and indirect impacts received by factor j from all other factors.By using the approach that was introduced by researchers of this paper, didinitial weighting respect to (15).

$$r_i = \sum_{1 \le j \le n} t_{ij} \qquad (13)$$

$$C_j = \sum_{1 \le i \le n} t_{ij} \qquad (14)$$

$$w = \frac{\sum\limits_{j=1}^{n} r_i + c_i}{\sum\limits_{i=1}^{n} \sum\limits_{j=1}^{n} r_i + c_i} \qquad (15)$$

## 3.3. TODIM

TODIM is a discrete multi-criteria method founded on prospect theory. The TODIM method has been successfully used and empirically validated in different applications. This is an experimental method based on how people make effective decisions in risky conditions. The shape of the value function of TODIM is identical to the prospect theory's gain and loss function. The global multi-criteria value function of TODIM then aggregates all measures of gains and losses by considering all criterion.

Fuzzy TODIM was used for supplier selection in green supply chain to select supplier with criteria of green supply chain management. According to Arshadi, Mahmoodi [2014],[16]the cause of advantage of TODIM can create weights for each criterion in supply chain respect to KM, and the evaluation is more exact than the other tools of MCDM. In this paper Fuzzy TODIM has been used that was introduced by Ming-Lang Tseng, KimHua Tan, Ru-Jen Lin, Yong Geng (2012)[17].

In this research, fuzzy TODIM has been improved with deffuzification method which has been used by Sachin K.Patil & Ravi Kant (2014).[8]The new TODIM improvement is in the first step of TODIM, where the weights of criteria are determined. With this improvement, defuzzification becomes more exact and the compliances of problem will be reduced.

The TODIM method uses paired comparisons between the criterions by using technically simple resources to eliminate occasional inconsistencies resulting from these comparisons. TODIM allows value judgments to be performed on a verbal scale using hierarchy of criteria, fuzzy value judgments and interdependence relationships among the alternatives. The decision matrix consists of alternatives and criterion. The alternatives $A_1,A_2,\ldots,A_m$ are viable alternatives, $c_1,c_2,\ldots,c_n$ are criterion, and $x_{ij}$ indicates the rating of alternative $A_i$ according to the criteria $c_j$. The weight vector w= $(w_1,w_2,\ldots w_n)$ comprises the individual weights $w_j(j=1,\ldots n)$ for each criterion $c_j$ satisfying $\sum^{n}_{i=1} w_j=1$. The data of decision matrix A originate from different sources. The matrix must be normalized to be dimensionless and allow various criterion to be compared to each other. In this study, the normalized decision matrix $R=[r_{ij}]_{m\times n}$ with i=1,…,m and j=1,…,n. 'A' matrix in this research is the same as 'A' matrix in DEMATEL which has been mentioned.

$$A=\begin{pmatrix} x_{11} & \cdots & x_{1n} \\ \vdots & \ddots & \vdots \\ x_{m1} & \cdots & x_{mn} \end{pmatrix} \qquad (10)$$

TODIM then calculates the partial dominance matrices and the final dominance matrix. The first calculation that the decision makers must define is a reference criteria (typically the criteria with the greatest importance weight). Therefore, $w_{rc}$ indicates the weight of the criteria c by the reference criteria r. These weights were determined by DEMATEL approach. TODIM is expressed by the following equations:

The dominance of each alternative over each alternative is:

$$\delta(A_i,A_j)=\sum_{c=1}^{m}\phi_c(A_i,A_j)_{\forall(i,j)} \qquad (16)$$

Where

$$\phi(A_i, A_j) = \begin{cases} \sqrt{\dfrac{w_{rc}(x_{ic} - x_{jc})}{\sum_{c=1}^{m} w_{rc}}} & if\,(x_{ic} - x_{jc}) > 0 \\ 0 & if\,(x_{ic} - x_{jc}) = 0 \\ \dfrac{-1}{\theta}\sqrt{\dfrac{(\sum_{c=1}^{m} w_{rc})(x_{ic} - x_{jc})}{w_{rc}}} & if\,(x_{ic} - x_{jc}) < 0 \end{cases} \quad (12)$$

The term $\varphi_c$ ($A_i$,$A_j$) represents the contribution of criterion c (c=1… m) to the function δ ($A_i$,$A_j$) when comparing alternative i with alternative j. The parameter θ represents the attention factor of the losses, whose mitigation depends on the specific problem. A positive ($x_{ic}$-$x_{jc}$) represents a gain. Whereas a nil or a negative ($x_{ic}$-$x_{jc}$) represents a loss. The final matrix of dominance is obtained by summing the partial matrices of dominance for each criterion. M.-L. Tseng et al.(2012),[17] The global value of the alternative I is determined by normalizing the final matrix of dominance according to the following expression:∑δ(i,j)

$$\xi_i = \frac{\sum_{j=1}^{n} \delta(i,j) - \min \sum_{j=1}^{n} \delta(i,j)}{\max \sum_{j=1}^{n} \delta(i,j) - \min \sum_{j=1}^{n} \delta(i,j)} \quad (17)$$

# 4. Methodology

## 4.1. Proposed approach

The method used to find out the weights of the KM criteria in supply chain management has introduced earlier and has been used in this paper is declared in the following. This is the first time of using the combination of TODIM and DEMATEL in fuzzy environment. In another words, the proposed approach in this research, introduces the combination procedure of fuzzy DEMATEL and TODIM. This is the innovation or contribution of this research.

Step 1: by studying papers and the literature review about KM criterion in supply chain and considering the relevant situation of supply chain in Iran, criterion of the problem are determined as shown in Table 1.

Step2: Ten experts which are professional in KM and supply chain network were selected and they were asked to fill the comparison matrix for criterion which were introduced in Table 1. (Experts were assistant Professors and PhD students and MS students. Their fields was industrial engineering, MBA, information system)

Step 3:Using deffuzifaction process which mentioned in the methodology,(2),(3),(4),(5),(6),(7),(8),(9) to determine initial matrix for DEMATEL approach.

Step4: Using Fuzzy DEMATEL for determining initial weights of criterion respect to the equation

Step5: Using initial weights and TODIM approach for the final ranking and weighting.

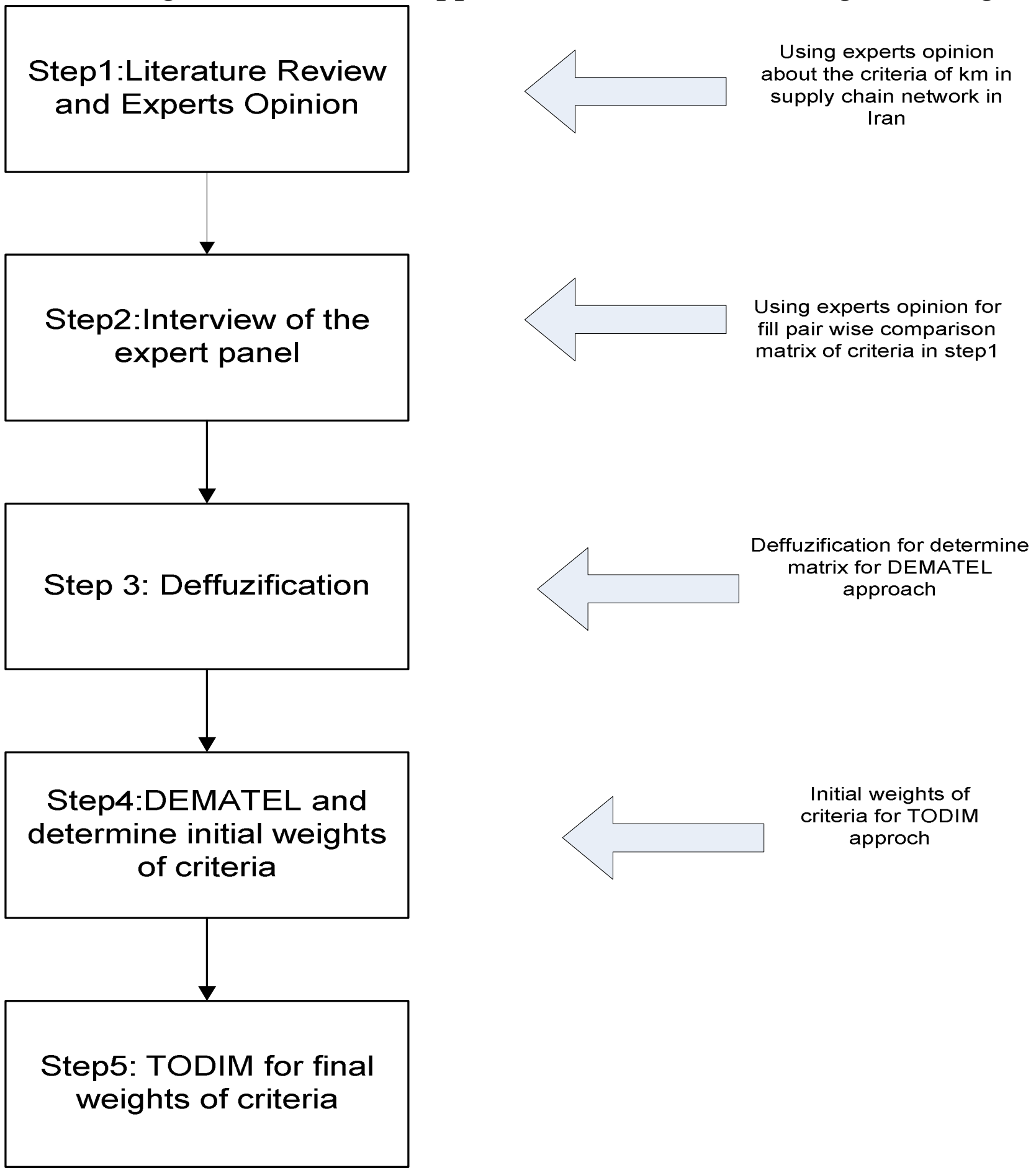

Figure 3. Flowchart of the Method

## 4.2. An illustrative example

Step1: First criterion were extracted from literature review and then by consulting with introduced experts, some of them became eliminated and some of them adjusted to Iran's supply chains and then after these efforts, all of the criterion categorized in this way that those similar criterion in context, located in the same category.

Step2. Ten experts were chosen in this research. They were professors of universities and PHD and MS students in industrial engineering and industrial management information systems and knowledge management. They all were those academic persons who not only were successful in academic activities, but also has worked in different industries and were familiar with both supply chain and knowledge management.

Step3: In this step experts opinions about criterion in format of the pair wise comparison matrix were collected. The A matrix was determined with using deffuzification approach which has been introduced in the section 3.2.Table 2 shows one expert opinions. The opinion of ten experts respect to (2) to (9) shown in Table 3.

| Criteria | 1 | 2 | 3 | 4 | 5 | 6 | 7 | 8 | 9 | 10 | 11 | 12 | 13 | 14 | 15 | 16 | 17 |
|---|---|---|---|---|---|---|---|---|---|---|---|---|---|---|---|---|---|
| 1 | | NO | NO | NO | VL | VL | VL | NO | NO | NO | L | VL | L | VL | VL | VL | NO |
| 2 | VH | | L | L | H | H | H | L | L | L | H | H | H | H | H | VH | H |
| 3 | VH | L | | L | L | L | H | L | L | L | H | H | VH | L | L | H | L |
| 4 | VH | L | L | | L | L | L | L | L | L | H | H | VH | H | L | H | L |
| 5 | H | VL | L | L | | L | L | VL | VL | VL | L | VL | H | H | L | VH | L |
| 6 | VH | VL | L | L | L | | VL | L | L | L | H | H | VH | L | H | VH | VL |
| 7 | H | VL | VL | L | L | H | | VL | VL | VL | L | H | L | NO | VL | VL | NO |
| 8 | VH | L | L | L | H | L | H | | L | H | VH | H | H | L | H | VH | L |
| 9 | VH | L | L | L | H | L | H | L | | VH | H | VH | H | L | L | H | L |
| 10 | VH | L | L | L | H | L | H | VL | NO | | H | L | L | VL | L | L | NO |
| 11 | L | VL | VL | VL | L | VL | L | NO | VL | VL | | VL | L | NO | VL | L | NO |
| 12 | H | VL | VL | VL | H | VL | VL | VL | NO | L | H | | VL | NO | VL | VL | NO |
| 13 | L | VL | NO | NO | VL | NO | L | VL | VL | L | L | H | | NO | VL | VL | NO |
| 14 | H | VL | L | VL | VL | L | VH | L | L | H | VH | VH | VH | | VH | VH | L |
| 15 | H | VL | L | L | L | VL | H | VL | L | L | H | H | H | VH | | L | NO |
| 16 | H | NO | VL | VL | NO | NO | H | NO | VL | L | L | H | H | NO | L | | NO |
| 17 | VH | VL | L | L | L | H | VL | L | L | VH | VH | VH | VH | L | VH | VH | |

Table 2. Expert opinion respect to Table 1.

| | C1 | C2 | C3 | C4 | C5 | C6 | C7 | C8 | C9 | C10 | C11 | C12 | C13 | C14 | C15 | C16 | C17 |
|---|---|---|---|---|---|---|---|---|---|---|---|---|---|---|---|---|---|
| C1 | | 0.395 | 0.11 | 0.031 | 0.156 | 0.287 | 0.098 | 0.11 | 0.121 | 0.086 | 0.086 | 0.25 | 0.305 | 0.375 | 0.287 | 0.287 | 0.273 |
| C2 | 0.226 | | 0.34 | 0.27 | 0.32 | 0.204 | 0.181 | 0.193 | 0.197 | 0.203 | 0.213 | 0.198 | 0.191 | 0.197 | 0.197 | 0.287 | 0.25 |
| C3 | 0.425 | 0.326 | | 0.315 | 0.305 | 0.273 | 0.203 | 0.25 | 0.25 | 0.395 | 0.395 | 0.213 | 0.402 | 0.392 | 0.303 | 0.298 | 0.463 |
| C4 | 0.455 | 0.311 | 0.315 | | 0.315 | 0.351 | 0.421 | 0.293 | 0.293 | 0.325 | 0.317 | 0.393 | 0.381 | 0.352 | 0.423 | 0.423 | 0.213 |
| C5 | 0.215 | 0.156 | 0.121 | 0.086 | | 0.121 | 0.098 | 0.086 | 0.086 | 0.211 | 0.318 | 0.215 | 0.312 | 0.314 | 0.217 | 0.198 | 0.098 |
| C6 | 0.483 | 0.185 | 0.295 | 0.311 | 0.375 | | 0.362 | 0.311 | 0.299 | 0.412 | 0.408 | 0.399 | 0.384 | 0.326 | 0.392 | 0.411 | 0.214 |
| C7 | 0.436 | 0.272 | 0.326 | 0.217 | 0.381 | 0.198 | | 0.215 | 0.198 | 0.295 | 0.306 | 0.276 | 0.205 | 0.198 | 0.386 | 0.295 | 0.196 |
| C8 | 0.452 | 0.295 | 0.317 | 0.311 | 0.433 | 0.14 | 0.364 | | 0.317 | 0.366 | 0.462 | 0.451 | 0.381 | 0.401 | 0.298 | 0.369 | 0.305 |
| C9 | 0.411 | 0.311 | 0.295 | 0.281 | 0.409 | 0.279 | 0.403 | 0.308 | | 0.371 | 0.418 | 0.385 | 0.433 | 0.375 | 0.271 | 0.38 | 0.291 |
| C10 | 0.425 | 0.311 | 0.291 | 0.304 | 0.363 | 0.217 | 0.276 | 0.184 | 0.195 | | 0.302 | 0.265 | 0.296 | 0.195 | 0.191 | 0.268 | 0.086 |
| C11 | 0.359 | 0.197 | 0.197 | 0.287 | 0.311 | 0.287 | 0.273 | 0.11 | 0.086 | 0.27 | | 0.287 | 0.273 | 0.11 | 0.25 | 0.273 | 0.11 |
| C12 | 0.411 | 0.195 | 0.183 | 0.214 | 0.365 | 0.197 | 0.286 | 0.098 | 0.163 | 0.294 | 0.311 | | 0.295 | 0.198 | 0.201 | 0.285 | 0.11 |
| C13 | 0.421 | 0.192 | 0.19 | 0.197 | 0.266 | 0.284 | 0.392 | 0.198 | 0.086 | 0.257 | 0.276 | 0.295 | | 0.086 | 0.191 | 0.295 | 0.086 |
| C14 | 0.42 | 0.275 | 0.28 | 0.301 | 0.375 | 0.298 | 0.392 | 0.195 | 0.185 | 0.365 | 0.421 | 0.364 | 0.401 | | 0.256 | 0.362 | 0.32 |
| C15 | 0.42 | 0.186 | 0.195 | 0.185 | 0.341 | 0.198 | 0.187 | 0.312 | 0.301 | 0.410 | 0.395 | 0.370 | 0.402 | 0.301 | | 0.440 | 0.230 |
| C16 | 0.451 | 0.201 | 0.195 | 0.186 | 0.321 | 0.195 | 0.301 | 0.140 | 0.193 | 0.274 | 0.265 | 0.298 | 0.310 | 0.198 | 0.185 | | 0.086 |
| C17 | 0.442 | 0.285 | 0.291 | 0.286 | 0.456 | 0.465 | 0.410 | 0.295 | 0.265 | 0.425 | 0.465 | 0.423 | 0.432 | 0.265 | 0.365 | 0.463 | |

Table 3. Represents deffuzification after combination opinions of ten experts.

In Table 3, experts opinions respect to deffuzification process were changed into crisp for DEMATEL process.

Step4: Using DEMATEL for determining initial weights of criteria.

To determine initial weights of criteria, DEMATEL approach used in this part .The matrix in Table 3 is decision matrix for DEMATEL. The indirect and direct weights and initial weights which is combination of direct and indirect weights shown in Table 4.

| | $r_{i(Direct)}$ | $C_{j(Indircet)}$ | $w_{initial}$ | Rank |
|---|---|---|---|---|
| C1 | 4.830062 | 2.760723 | 0.07955 | 1 |
| C2 | 3.489151 | 1.810724 | 0.055542 | 11 |
| C3 | 3.340061 | 1.686741 | 0.05268 | 13 |
| C4 | 3.242659 | 1.615606 | 0.050914 | 14 |
| C5 | 4.244822 | 2.334578 | 0.068951 | 2 |
| C6 | 3.500196 | 1.809355 | 0.055643 | 10 |
| C7 | 3.747295 | 1.976808 | 0.059988 | 8 |
| C8 | 2.95099 | 1.404065 | 0.04564 | 16 |
| C9 | 2.914474 | 1.378189 | 0.044986 | 17 |
| C10 | 3.911771 | 2.089776 | 0.062895 | 7 |
| C11 | 4.135935 | 2.247816 | 0.066901 | 4 |
| C12 | 3.962638 | 2.136897 | 0.063922 | 6 |
| C13 | 4.210022 | 2.314495 | 0.068376 | 3 |
| C14 | 3.559882 | 1.849966 | 0.056694 | 9 |
| C15 | 3.438394 | 1.773175 | 0.054616 | 12 |
| C16 | 4.106032 | 2.238612 | 0.066491 | 5 |
| C17 | 2.98092 | 1.428692 | 0.046212 | 15 |

Table 4. DEMATEL approach and initial weights.

Step5: Using TODIM with respect weights in step 4 and final ranking of criteria and determining final weights: after determining the initial weights, the TODIM approach used to determine final weights of criteria of KM in supply chain respect to each other .Table 5. The final weights columns shows the normalize weight in TODIM approach, and the $\xi$ columns shows distance from gain and losses.

| | $\xi_i$ | Final weights | Hybrid Ranking | DEMATEL Ranking |
|---|---|---|---|---|
| C1 | 0.0287 | 0.607 | 2 | 1 |
| C2 | 0.0325 | 0.0605 | 4 | 11 |
| C3 | 0.0803 | 0.0575 | 12 | 13 |
| C4 | 0.0751 | 0.0578 | 11 | 14 |
| C5 | 0.0039 | 0.0628 | 1 | 2 |
| C6 | 0.0927 | 0.0567 | 14 | 10 |
| C7 | 0.0594 | 0.0588 | 9 | 8 |
| C8 | 0.1007 | 0.0562 | 16 | 16 |
| C9 | 0.0947 | 0.0566 | 15 | 17 |
| C10 | 0.0516 | 0.0593 | 8 | 7 |
| C11 | 0.0314 | 0.0605 | 5 | 4 |
| C12 | 0.0354 | 0.0603 | 6 | 6 |
| C13 | 0.0306 | 0.0606 | 3 | 3 |
| C14 | 0.0860 | 0.0572 | 13 | 9 |
| C15 | 0.0693 | 0.0582 | 10 | 12 |
| C16 | 0.0367 | 0.0602 | 7 | 5 |
| C17 | 0.100 | 0.0562 | 17 | 15 |

Table 5. TODIM approach for determining final weights

# 5. Conclusion

In recent years the effect of KM in supply chain in Iran has had an important role, and the goal of this study is to find a solution for this problem. The hybrid TODIM-DEMATEL fuzzy approach effectively shows the importance of supporting experiment in supply chain in Iran, which means for improving supply chain is really important to start from experiment to avoid cost and other restriction of resources.

Results show that the KM indices according to their importance are as below:

Virtual teaming, supporting experimentation in the SC, networking techniques, electronic data interchange (EDI) links, Creating a compatible communication/information system, IT integration with all suppliers/customers, Transferring knowledge quickly and efficiently throughout the organization, Systematic problem-solving, Integrated business systems, Giving reward to employees for their new ideas in the supply chain, Capacity to develop knowledge within SC, Establishing more frequent contact with supply-chain members, Employee involvement, Developing new insights, Employee empowerment, Learning from their own experiences and past history, Learning from the experiences and best practices of others.

So the hypothesis can be proved because all of the criterion in table 1, are said to be key criterion of KM adaption in supply chain. By paying attention to each of them respectively, a firm can successfully implement a KM program through its supply chain.